\documentclass[twocolumn,pra,showpacs,preprintnumbers,superscriptaddress]{revtex4}
\usepackage{amsmath,multirow,slashbox,amssymb,graphicx,graphics,lipsum}
\usepackage{times}

\begin{document}


\title{\textbf{Detecting topological exceptional points in a parity-time symmetric system with cold atoms}}

\author{Jian Xu}
\email{xujian_328@163.com}
\affiliation{College of Electronics and Information Engineering, Guangdong Ocean University, Zhanjiang 524088,  China}
\affiliation{Guangdong Provincial Key Laboratory of Quantum Engineering and Quantum Materials,
SPTE, South China Normal University, Guangzhou 510006, China}

\author{Yan-Xiong Du}
\affiliation{Guangdong Provincial Key Laboratory of Quantum Engineering and Quantum Materials,
SPTE, South China Normal University, Guangzhou 510006, China}

\author{Wei Huang}
\affiliation{Guangdong Provincial Key Laboratory of Quantum Engineering and Quantum Materials,
SPTE, South China Normal University, Guangzhou 510006, China}

\author{Dan-Wei Zhang}
\email{zdanwei@126.com}
\affiliation{Guangdong Provincial Key Laboratory of Quantum Engineering and Quantum Materials,
SPTE, South China Normal University, Guangzhou 510006, China}

\date{\today}

\begin{abstract}
We reveal a novel topological property of the exceptional points in a
two-level parity-time symmetric system and then propose a scheme to detect
the topological exceptional points in the system, which is embedded in a larger Hilbert space
constructed by a four-level cold atomic system. We show that a tunable
parameter in the presented system for simulating the non-Hermitian
Hamiltonian can be tuned to swept the eigenstates through the
exceptional points in parameter space. The non-trivial Berry
phases of the eigenstates obtained in this loop from the
exceptional points can be measured by the atomic
interferometry. Since the proposed operations and detection are
experimentally feasible, our scheme may pave a promising way to
explore the novel properties of non-Hermitian systems.
\end{abstract}


\maketitle


\section{INTRODUCTION}

Non-Hermitian Hamiltonian used to describe open or dissipative
systems usually have complex eigenvalues. However, it is recently
found that a series of non-Hermitian Hamiltonians have real
eigenvalues if they are invariant under the parity $(P)$ and
time-reversal $(T)$ union operation \cite{C. M. Bender1,C. M.
Bender2,C. M. Bender3}. Meanwhile, the eigenstates of these
Hamiltonians are also commuted with the PT-symmetry
operation\cite{T. Kottos}. Due to various intriguing properties, the
PT-symmetric systems have raised broad attentions. Some PT-symmetric models have been
experimentally realized in physical systems, such as active
LRC circuits \cite{J. Schindler,Z. Lin}, two coupled waveguides
\cite{C. E. Ruter}, photonic lattices \cite{A. Szameit,A.
Regensburger}, microwave billiard \cite{S. Bittner}, transmission
line \cite{Y. Sun}, whispering-gallery microcavities \cite{L.
Chang,B. Peng} and single-mode lasing \cite{L. Feng,H. Hodaei}.
Recently, the schemes for simulating the PT-symmetry potentials with cold atoms in optical systems have been theoretically proposed \cite{C.
Hang1,J.-H. Wu,C. Hang2} and experimentally realized \cite{P. Peng}.

Exceptional point (EP) is a special point in parameter space of
non-Hermitian systems where both eigenvalues and eigenstates
coalesce into only one value and state \cite{T. Kato,H. Cao}. One
of the important properties of EPs is that the states of EPs are
chiral \cite{W.D. Heiss1}, which has been experimentally observed
\cite{C. Dembowski1,Y. Choi}. On the other hand, the chirality
leads to an unique effect that the eigenstates exchange themselves
but only one of them obtains a $\pi$ Berry phase in a cyclic
evolution \cite{W.D. Heiss2,W.D. Heiss3}. This chiral phenomenon
of EPs has been experimentally demonstrated in microwave cavity
for the first time \cite{C. Dembowski2,C. Dembowski3}. Very
recently, a full dynamical encircling of EP has been realized
\cite{J. Doppler} and a non-reciprocal topological energy transfer
due to dynamical encircling of such point has been measurement
\cite{H. Xu}. In contrast, other theoretical and numerical results
suggest in this case the eigenstates change to the other one but
both of them obtain a $\pi/2$ Berry phase due to the linear
dependence of eigenstates \cite{I. Rotter,H. Eleuch}, which has
been verified in a recent experiment \cite{B. Wahlstrand}.

In this paper, we demonstrate that there are two different chiral
EPs in parameter space and the non-trivial Berry phase of EPs
emerge different due to the chirality breaking when the
eigenstates exactly to sweep through EPs in a cyclic evolution.
Then we propose a scheme to realize the PT-symmetric Hamiltonian
in cold atomic systems that the parameters can be exactly
controlled in time. Based on the idea of the Naimark-Dilated
operation \cite{Uwe Gunther} and the embedding quantum simulator
\cite{X. Zhang}, we show that a two-level PT-symmetric Hamiltonian
can be constructed through a four-level Hermitian Hamiltonian in
an embedding cold atomic system. Finally, we propose to detect
this Berry phase through the observation of the phase difference
between atomic levels, which can be measured through atomic
interferometry. The proposed scheme provides a promising approach
to realize the PT-symmetric Hamiltonian in cold atomic systems and
to further explore the exotic properties of the EPs.

The paper is organized as follows. Section II describes our two-level PT-Hamiltonian and the topological properties of the intrinsic EPs. In Sec. III, we propose an experimentally feasible scheme for emulating the non-Hermitian two-level Hamiltonian in a Hermitian four-level cold atomic system. In Sec. IV, we show that the topological EPs can be measured by the atomic interferometry in the cold atom system. Finally, a brief discussion and a short
conclusion are given in Sec. V.

\section{topological exceptional points in a PT-symmetric system}

If a Hamiltonian $H$ is non-Hermitian for describing an open or
dissipative system, there are gain or loss effects in this system
and the eigenvalues are generally complex values. However, if
gain and loss of this system are balanced, this system remains
stable and all eigenvalues are real numbers. This phenomenon is
described by the PT-symmetric theory. Supposing that in this case
$\sigma_i$ $(i=x,y,z)$ are Pauli matrices, the parity operator $P$ is
$\sigma_x$ and the time-reversal operator $T$ is the complex
conjugation operator, which is an antilinear operator. A simple
PT-symmetric Hamiltonian can be constructed as
\begin{equation}
\label{two-level}
H_{PT}=S\begin{pmatrix} i\sin(\alpha) & 1 \\ 1 &  -i\sin(\alpha) \end{pmatrix},
\end{equation}
where $S$ is a general scaling factor of the matrix. The angle
$\alpha$ characterizes the non-Hermiticity of the Hamiltonian.
When $\alpha=0$, the Hamiltonian $H$ is a Hermitian operator, when
$\alpha\neq0$ the Hamiltonian $H$ becomes a non-Hermitian
operator. In the case of $\alpha=\pm\pi/2$, the eigenvalues and
the eigenstates coalesce into a single value and state,
respectively.

The eigenvalues of Eq. (\ref{two-level}) are $E_{\pm}(\alpha)=\pm\chi=\pm S\cos(\alpha)$ and the corresponding eigenstates are given by
\begin{equation}
\begin{split}
\label{eigenstates}
|E_+(\alpha)\rangle=\frac{e^{i\alpha/2}}{\sqrt{2\cos(\alpha)}}\begin{pmatrix} 1 \\   e^{-i\alpha} \end{pmatrix}, \\
|E_-(\alpha)\rangle=\frac{ie^{-i\alpha/2}}{\sqrt{2\cos(\alpha)}}\begin{pmatrix} 1 \\   -e^{i\alpha} \end{pmatrix}.
\end{split}
\end{equation}
In addition, the non-Hermitian Hamiltonian $H_{PT}$ has a bi-orthogonal basis ${|E_\pm(\alpha)\rangle},{|\Lambda_\pm(\alpha)\rangle}$ \cite{P. M. Morse}:
\begin{equation}
\begin{split}
\label{bi-orthogonal basis}
H_{PT}^*{|\Lambda_\pm(\alpha)\rangle}&=E_{\pm}^*(\alpha){|\Lambda_\pm(\alpha)\rangle},\\
{|\Lambda_\pm(\alpha)\rangle}&=-{|E_\mp(\alpha)\rangle}.\\
\end{split}
\end{equation}
When $\alpha=\pm\pi/2$, the two eigenvalues become $E_{\pm}=0$ and the corresponding eigenstates coalescence at the same time:
\begin{equation}
\begin{split}
\label{EPs eigenstates}
|E(\pi/2)\rangle\propto\begin{pmatrix} 1 \\   -i \end{pmatrix}, \\
|E(-\pi/2)\rangle\propto\begin{pmatrix} 1 \\   i \end{pmatrix}.
\end{split}
\end{equation}
The signs before $i$ in Eq. (\ref{EPs eigenstates}) depend on the system and give the chirality of these specific degeneracy points dubbed as EPs \cite{T. Kato, W.D. Heiss1,W.D. Heiss2,W.D. Heiss3}, such that the two EPs are different from each other. The Hamiltonian thus is restricted to purely real eigenvalues, and the time evolution operator $\hat{U}_{PT}(t)=e^{-iH_{PT}t}$ is unitary with the explicit form
\begin{equation}
\label{U two-level}
\hat{U}_{PT}(t)=
\frac{1}{\cos(\alpha)}\begin{pmatrix} \cos(\chi t-\alpha) & -i\sin(\chi t) \\ -i\sin(\chi t) &  \cos(\chi t+\alpha) ) \end{pmatrix}.
\end{equation}

With the analytical solution of the PT-symmetric Hamiltonian, we
can study the topological properties of the EPs. Considering the eigenvalues $E_{\pm}(\alpha)$, we can find that the eigenstates
$|E_+(\alpha)\rangle$ and $|E_-(\alpha)\rangle$ respectively represent the
higher and lower levels when $\alpha\in[-\pi/2+2N\pi,\pi/2+2N\pi]$, but respectively represent the lower
and higher levels when $\alpha\in[\pi/2+2N\pi,3\pi/2+2N\pi]$, with $N$ being a positive integer. This means that
the definition of the domain in the system is $[-\pi/2,\pi/2]$
and the eigenstates exchange themselves when $\alpha$ sweeps
through the EPs every time. With the eigenvalues $E_{\pm}(\alpha)=\pm S\cos(\alpha)$, one can also find that the
corresponding point of $\alpha$ in the other Riemann surface is
the point of $\pm\pi-\alpha$. Due to the degeneracy of the EPs,
the eigenstates obtain non-Abelian geometric phases through
passing the EPs. For this degenerate non-Hermitian system, the Berry
phase in the cyclic evolution is \cite{S. W. Kim}
\begin{eqnarray}
\label{Berry phase}
\gamma=\oint_C A d\alpha,
\end{eqnarray}
where $A$ is the non-Abelian Berry connection:
\begin{equation}
\label{Berry connection}
A=
i\begin{pmatrix} \langle \Lambda_+|d_\alpha|E_+\rangle & \langle \Lambda_+|d_\alpha|E_-\rangle  \\ \langle \Lambda_-|d_\alpha|E_+\rangle & \langle \Lambda_-|d_\alpha|E_-\rangle \end{pmatrix},
\end{equation}
with $d_\alpha$ being the $\alpha$ derivative. For the PT-symmetric Hamiltonian $H_{PT}$ here, we can obtain the non-Abelian geometric phases for two different loops from $\alpha$ to $\pm\pi-\alpha$ (which pass through two EPs of different chiralities) as
\begin{eqnarray}
\label{Berry phase2}
\gamma_{\alpha\rightarrow\pm\pi-\alpha}=\begin{pmatrix} 0 & \pm \frac{\pi}{2}  \\ \pm \frac{\pi}{2} & 0 \end{pmatrix}.
\end{eqnarray}
Consequently, when $\alpha$ successively sweeps through two
different EPs in the same evolutionary direction, the eigenstates
become original with an additional $\pi$ Berry phase. In this case, the
eigenstates under the whole evolution can be written as
\begin{equation}
\begin{split}
\label{pass through EP2}
&|E_\pm(\alpha)\rangle {\longrightarrow} e^{i\frac{\pi}{2}}|E_\mp(\pi-\alpha)\rangle {\longrightarrow}-|E_\pm(\alpha)\rangle, \\
&|E_\pm(\alpha)\rangle {\longrightarrow} e^{-i\frac{\pi}{2}}|E_\mp(-\pi-\alpha)\rangle {\longrightarrow}-|E_\pm(\alpha)\rangle.
\end{split}
\end{equation}
Unlike $|E_\pm\rangle {\rightarrow} |E_\mp\rangle {\rightarrow}-|E_\pm\rangle$ or $|E_\pm\rangle {\rightarrow} -|E_\mp\rangle {\rightarrow}-|E_\pm\rangle$ in the general case, here Eq. (\ref{pass through EP2}) shows that there is an obviously different behavior in the intermediate process when one passes through different EPs successively.

The above results show that in non-Hermitian systems, the
eigenvalue surfaces exhibit a complex-square-root  topology with a
branch point named by EP, which can also be considered as a
critical point where a transition from PT-symmetric phase to
broken-PT-symmetric phase. A consequence of this topology is that
encircling an EP once in the parameter space results in the
exchange of both eigenvalues and eigenstates. This means that one
has to encircle an EP twice to recover the original eigenvalues
and eigenstates. On the other hands, one of the eigenstates acquires a
Berry phase of $\pm\pi$ when encircling an EP once and the other
one acquires the same phase in the second loop. However, because one not longer distinguishes the
clockwise or anticlockwise direction of the state evolution, the
result of passing through the EPs once will be different. In this system, one
acquires this phase in each loop is determined by the chirality of
the EP and the evolutionary direction of the eigenstates in the
parameter space.

\section{realization of the two-level PT-symmetric Hamiltonian in a four-level cold atomic system}

The PT-symmetric Hamiltonian may be difficult to achieve in
a practical non-Hermitian two-level system. In this section, we propose to use a
four-level Hermitian system to simulate the two-level PT-symmetric
Hamiltonian (1). In the basis
$(|1\rangle,|2\rangle,|3\rangle,|4\rangle)^T$, the four-level
Hermitian Hamiltonian takes the form
\begin{equation}
\begin{split}
\label{four level H 3}
H_F=\chi\begin{pmatrix} 0 & \cos(\alpha) & i \sin(\alpha) & 0 \\ \cos(\alpha) & 0 & 0 & -i \sin(\alpha)\\ -i \sin(\alpha) & 0 & 0 & \cos(\alpha)\\ 0 & i \sin(\alpha) &\cos(\alpha) & 0 \end{pmatrix}.\\
\end{split}
\end{equation}
The corresponding time evolution operator is given by $\hat{U}_F=e^{-iH_Ft}$. For an arbitrary state $\upsilon=(a,b)^T$, we can find that:
\begin{equation}
\begin{split}
\label{four level U}
\hat{U}_F \begin{pmatrix} \upsilon \\ \eta \upsilon \\\end{pmatrix}=
\begin{pmatrix} \hat{U}_{PT}(t) & 0 \\ 0 & \eta \hat{U}_{PT}(t) \eta^{-1} \end{pmatrix} \begin{pmatrix} \upsilon \\ \eta \upsilon \\\end{pmatrix},
\end{split}
\end{equation}
with $\eta=\begin{pmatrix} 1 & -i \sin (\alpha) \\ i\sin(\alpha) & 1 \end{pmatrix}/\cos(\alpha)$. So if we take the four-level states as $(|E_\pm(\alpha)\rangle,\eta |E_\pm(\alpha)\rangle)^T$, we can simulate the evolution of the states in Eq. (\ref{eigenstates}) in the two-level PT-symmetric Hamiltonian.

\begin{figure}[tbp]
\includegraphics[width=6.5cm]{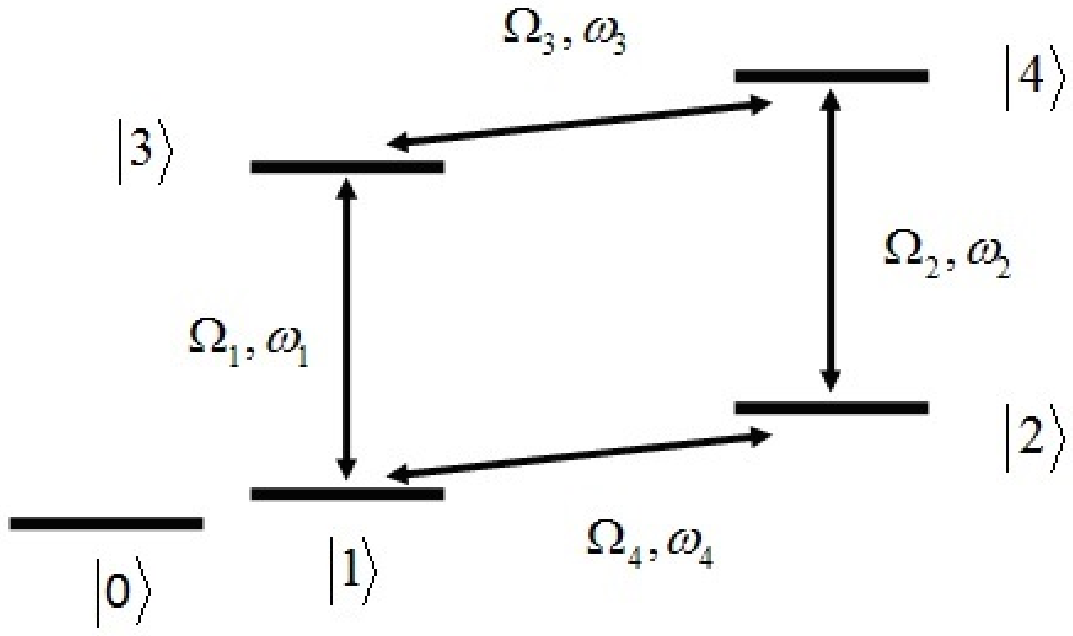}
\label{fig1} \caption{Schematic representation of the light-atom
interaction configuration of the four-level Hamiltonian. The
relevant atomic levels are coupled by microwave radios with
corresponding Rabi frequencies.}
\end{figure}

Now we present an experimentally realizable scheme implementing with cold atoms. We consider an atomic cloud of $^{87}$Rb with five
internal states in the ground-state manifold,
noted by $|1+m_F\rangle=|F=1,m_F=-1,0,1\rangle$ and
$|4+m_F\rangle=|F=2,m_F=-1,0\rangle$, which are separated by the
hyperfine splitting $\omega_{HF}$ and the Zeeman splitting
$\omega_Z$ caused by a uniform static magnetic field. One can apply four microwave radios
to couple these atomic levels through the annulus configuration as shown in Fig.
1. Here $\Omega_i$ and $\omega_i$ denote the Rabi frequencies and the
frequencies of the corresponding microwave radios, respectively. In
particular, we use resonant microwaves to couple
$|1\rangle\longleftrightarrow|3\rangle$ and
$|2\rangle\longleftrightarrow|4\rangle$ and use radio-frequency
fields to couple $|1\rangle\longleftrightarrow|2\rangle$ and
$|3\rangle\longleftrightarrow|4\rangle$, respectively. Supposing that $\omega_{ei}$ are the energies of $|i\rangle$ and the energy of
$|1\rangle$ is the zero of energy, the total Hamiltonian can be written as $H=H_0+H_{int}$ with
\begin{equation}
\begin{split} \label{H four-level}
H_0=&\sum_{j}(\omega_{ej}-\omega_{e1})|j\rangle\langle j|,\\
H_{int}=&\Omega_1e^{i\omega_1t}|3\rangle\langle1|+\Omega_2e^{i\omega_2t}|3\rangle\langle4|+\\
\Omega_3e^{i\omega_3t}&|4\rangle\langle3|+\Omega_4e^{i\omega_3t}|2\rangle\langle1|+H.c.,
\end{split}
\end{equation}
In the bare-state basis
$(|1\rangle,|2\rangle,|3\rangle,|4\rangle)^T$, one has
\begin{equation}
\label{transformation matrix}
V=
\begin{pmatrix} 1 & 0 & 0 & 0 \\ 0 & e^{-\omega_4t} & 0 & 0\\ 0 & 0 & e^{-\omega_1t} & 0\\ 0 & 0 & 0 & e^{-(\omega_1+\omega_3)t} \end{pmatrix}
\end{equation}
and the Hamiltonian in the rotating frame becomes
\begin{widetext}
\begin{equation}
\label{four level H 2}
\tilde{H}=
i\frac{dV^{\dag}}{dt}V+V^{\dag}HV
=\begin{pmatrix} 0 & \Omega_4 & \Omega_1 & 0 \\ \Omega_4^* & -\omega_4-\omega_{e1}+\omega_{e2} & 0 & e^{-i(\omega_1+\omega_2-\omega_3-\omega_4)t}\Omega_3\\ \Omega_1^* & 0 & -\omega_1-\omega_{e1}+\omega_{e3} & \Omega_2\\ 0 & e^{i(\omega_1+\omega_2-\omega_3-\omega_4)t}\Omega_3^* & \Omega_2^* & -\omega_1-\omega_3-\omega_{e1}+\omega_{e4} \end{pmatrix}.
\end{equation}
\end{widetext}
To ensure the Hamiltonian being time-independent, we choose
$\omega_1+\omega_2=\omega_3+\omega_4$. On the other hand,
considering the resonance condition
$\Delta_1=\omega_{e3}-\omega_{e1}-\omega_1=0$,
$\Delta_2=\omega_{e4}-\omega_{e3}-\omega_2=0$,
$\Delta_3=\omega_{e4}-\omega_{e2}-\omega_3=0$,
$\Delta_4=\omega_{e2}-\omega_{e1}-\omega_4=0$, with
$\omega_1=\omega_2=\omega_{HF},\omega_3=\omega_4=\omega_{Z}$.
In particular, we choose the corresponding Rabi frequencies
$\Omega_1=-\Omega_3=iS\sin(\alpha)
\cos(\alpha)$ and $\Omega_2=\Omega_4=S
\cos^2(\alpha)$. Under these conditions, the Hamiltonian $\tilde{H}$ can be
represented as $H_F$ in Eq. (\ref{four level H 3}):
\begin{equation}
\label{four level H 4}
\tilde{H}
=\begin{pmatrix} 0 & \Omega_4 & \Omega_1 & 0 \\ \Omega_4^* & 0 & 0 & \Omega_3\\ \Omega_1^* & 0 & 0 & \Omega_2\\ 0 & \Omega_3^* & \Omega_2^* & 0 \end{pmatrix}
=H_{F}.
\end{equation}
Up to this step, we have proposed a method to realize the required
four-level Hamiltonian for simulating the two-level PT-symmetric
Hamiltonian in a cold atomic system. It is noteworthy that in this
system, we can precisely and easily control the non-Hermitian
parameter $\alpha$ by adjusting $\Omega_i$.

\section{detecting the Berry phase in the PT-symmetric system}

In the section, we show how to detect the Berry phases of the mimicked EPs
in the four-level cold atomic system. First, we needs an additional atomic level for this
measurement, which is denoted by $|0\rangle$ in the cold atomic system as shown in Fig. 1. We assume the atoms
are initially pumped to $|0\rangle$ and the transitions
$|0\rangle\rightarrow|i\rangle (i=1,2,3,4)$ can be realized
successively through the stimulated-Raman-adiabatic-passage
(STIRAP) \cite{Kral,Du}. It is noted that the microwave radios must be
phase-locking between each STIRAP for keeping the coherence between the states. On the other hands, only the phase
difference between $|0\rangle$ and $|i\rangle$ is needed to be
concerned, so it is nonsignificant what the population differences
between $|0\rangle$ and other levels are. Under this condiction, we make
$|0\rangle\rightarrow|0\rangle+\eta|\psi_\pm(\alpha_0)\rangle^T$
from the very beginning, where $\eta$ is an arbitrary real number,
for the preparation of the PT-symmetric initial state. The phase difference between $|1\rangle$ and $|2\rangle$ is
the only distinction between $|E_{\pm}(\alpha)\rangle$ for a given
$\alpha$, and the phase differences between $|0\rangle$ and
$|1\rangle$ ($|2\rangle$) can be used to detect the Berry phases of
$|E_{\pm}(\alpha)\rangle$. Thus we can detected the Berry phases
by measuring the phase differences between the corresponding atomic levels.

\begin{figure}[tbp]
\includegraphics[width=8cm]{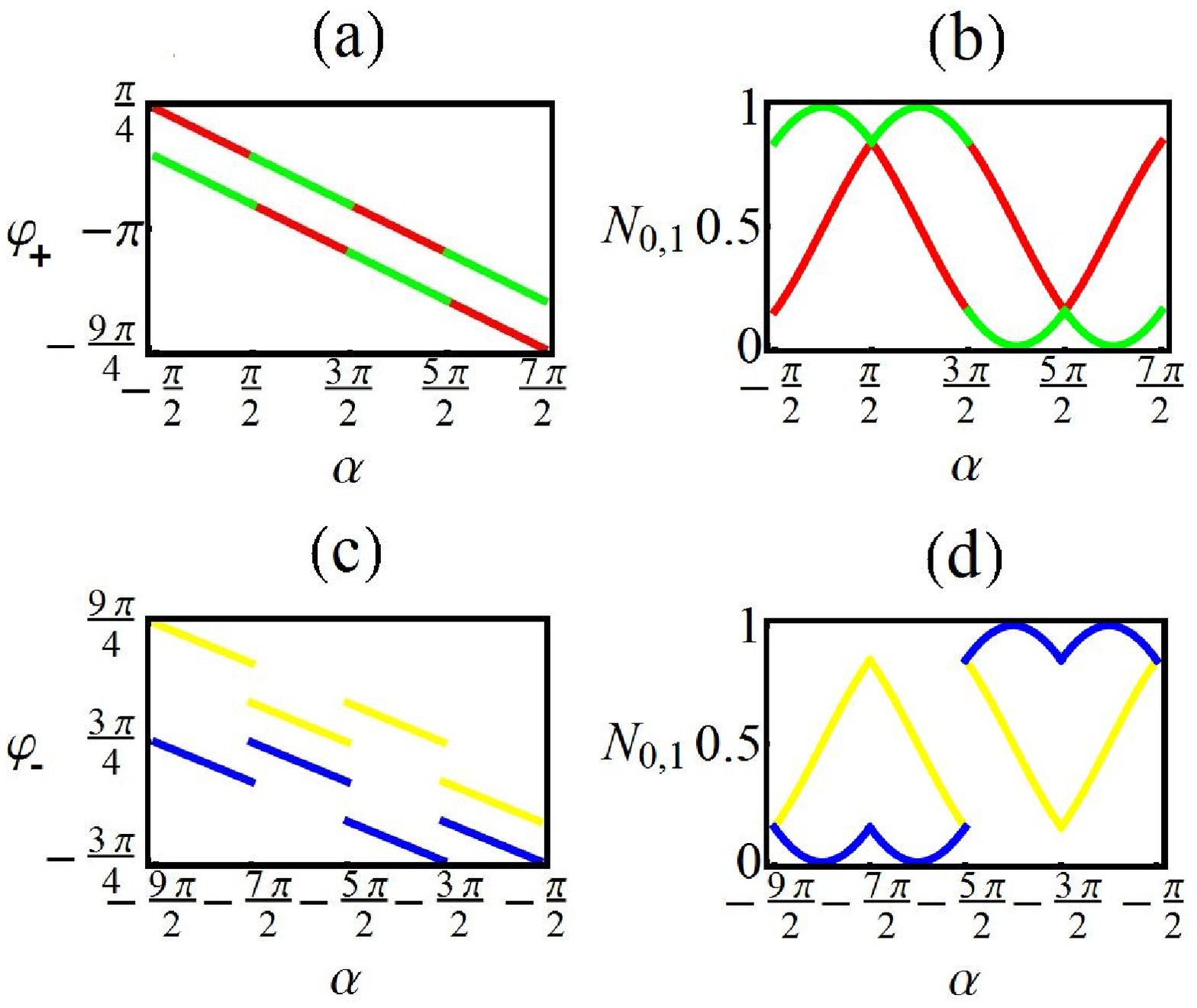}
\label{fig2} \caption{(Color online) The phase difference and
atomic population versus angle. The red, green, yellow and blue
line denote  the $|E_+(\alpha)\rangle$, $|E_-(\pi-\alpha)\rangle$,
$|E_-(\alpha)\rangle$ and $|E_+(\pi-\alpha)\rangle$, respectively.
(a,c) The dynamic phase $\gamma_d=0$. (b,d) The dynamic phase
$\gamma_d=\pi/2$.}
\end{figure}

The phase difference between two atomic levels can be measured
through the atomic interferometry. For an arbitrary state denoted by
$|a\rangle+e^{-i \varphi}|b\rangle$, a $\pi/2$-pulse operation takes the form
\begin{equation}
\label{pi/2}
U_{\pi/2,\phi}=\left(   \begin{array}{cc}
1 & -i e^{-i \phi}\\
-i e^{i \phi} & 1\\
\end{array}
\right),
\end{equation}
with $\phi$ being a controllable phase of the $\pi/2$-pulse. After
applying a $\pi/2$-pulse to the state $|a\rangle+e^{-i \varphi}|b\rangle$,
one can find the relationship between the atomic populations and
the phase differences as
\begin{equation}
\label{N1}
N_{a,b}=\frac{1}{2}\left[1\mp\sin(\phi+\varphi)\right].
\end{equation}
Considering Eq. (\ref{eigenstates}) and Eq. (\ref{N1}), the atomic populations of the different levels are given by
\begin{eqnarray}
\label{N2}
N_{1,2}(\alpha)=
\begin{cases}
\left[1\mp\sin(\phi+\alpha)\right]/2 & ~\text{for}~~  |E_{+}(\alpha)\rangle\\
\left[1\mp\sin(\phi+\pi-\alpha)\right]/2 & ~\text{for}~~  |E_{-}(\alpha)\rangle\\
\end{cases}.
\end{eqnarray}
For a given $\alpha$, the function of $\phi$ in Eq. (\ref{N2}) can
be used to distinguish $|E_{+}(\alpha)\rangle$ and $|E_{-}(\alpha)\rangle$, so we
can verify whether the eigenstates exchange themselves
through measuring the function of $\phi$ in Eq. (\ref{N2}) when sweeping an EP.

After confirming the two states $|E_{\pm}(\alpha)\rangle$, we can measure
the phase difference between $|0\rangle$ and $|1\rangle$ to
determine the Berry phases $\gamma_\pm$ for
$|E_{\pm}(\alpha)\rangle$, which is related to the value of
$\alpha$ and the evolution loop. For
$|0\rangle$ and $|1\rangle$, we can also find the relationship
between atomic populations and total phase differences:
\begin{eqnarray}
\label{N3}
\begin{split}
&N_{0,1}(\alpha)=
\begin{cases}
\left(1\mp\sin\varphi_+\right) /2 & ~\text{for}~~  |E_{+}(\alpha)\rangle\\
\left(1\mp\sin\varphi_-\right) /2 & ~\text{for}~~  |E_{-}(\alpha)\rangle
\end{cases}
\end{split},
\end{eqnarray}
where the total phases between the corresponding atomic levels in the case of different eigenstates $\varphi_+=\gamma_d-\alpha/2-\theta_0(\alpha)\pi/2$ and $\varphi_-=\gamma_d-(\pi-\alpha)/2-\theta_0(\alpha)\pi/2$ with $\gamma_d=\mp S\cos(\alpha)t+\phi+\omega_Z t$ being the dynamic phase due to the evolution $\hat{U}_F$, the $\pi/2$-pulse and the Zeeman energy difference, respectively, and $\theta_0(\alpha)=\theta[-\cos(\alpha)]$ being a Heaviside unit step function whose value is 0 for $-\cos(\alpha)<0$ and 1 for $-\cos(\alpha)>0$. It is clear that the topology of the EP is the source of the Heaviside unit step function. Due to the symmetry of the trigonometric functions, confirming $|E_{+}(\alpha)\rangle$ and $|E_{-}(\alpha)\rangle$ in Eq. (\ref{N3}) needs two values of $\gamma_d$.

For simplicity, here we choose the dynamic phase as $0$ and $\pi/2$, and the phases $\varphi_\pm$ and the atomic populations $N_{0,1}$ in the different eigenstates are shown in Fig. (2). For a given eigenstate, one can measure the phases $\varphi_\pm$ to obtain the Berry phases of the eigenstates from Eq. (\ref{N3}). In particular, for a given eigenstate and the parameter $\alpha$,
we first measure $N_{1,2}$ in Eq. (\ref{N2}) to determine which
eigenstate is  through the function of $\phi$. Then one can detect the
total phase $\varphi_{\pm}$ by measuring $N_{0,1}$ in Eq.
(\ref{N3}), as shown in Fig. 2. After that, one can
control the systemic parameter $\alpha$ to sweep an EP in parameter
space that the eigenstate is predicted to obtain a non-Abelian
phase in Eq. (\ref{Berry phase2}). Again one can successively measure
$N_{1,2}$ and $N_{0,1}$ to determine whether the eigenstate has
been changed and the variation of total phase. Here we can find
that: i) The eigenstates obtain a phase of $+\pi/2$ or $-\pi/2$
alternately when $\alpha$ sweeps through the different EPs
successively. ii) The eigenstates exchange themselves when
$\alpha$ sweeps through an EP and the phase differences from
$|E_\pm(\alpha)\rangle$ to $|E_\mp(\pi-\alpha)\rangle$ are always
$\pm\pi/2$. iii) The eigenstates obtain a Berry phase of $\pm\pi$ when
$\alpha$ sweeps $\pm2\pi$. In short, the measurement of the phase
differences between $|i\rangle$ $(i=0,1,2)$ provides a simple way to
experimentally verify Eq. (\ref{pass through EP2}) and demonstrate the Riemann sheet structure and the intrinsic properties of the EPs.

\section{Discussion and Summary}

In the above case, there is no dynamical phase contribution between the four states $|i\rangle$ ($i=1,2,3,4$), but the evolution of $|i\rangle$ must be still adiabatic in order to avoid non-adiabatic transitions. To be specific, the four Rabi frequencies should keep the adiabatic cyclic evolutions in the parameter space with the change rate of $\alpha$ being significantly smaller than the typical Rabi frequencies. This means that the evolution period of the system is much larger than the inverses of the energy gaps between the atomic levels. On the other hands, with the non-adiabatic transition between $|E_{\pm}(\alpha)\rangle$ being avoided, the change rate of $\alpha$ is also smaller than $2\chi$, which is the difference of eigenvalue. Thus, the adiabatic condition takes the form
\begin{equation}
\label{adiabatic condition}
\frac{d \alpha}{d t}\ll \Omega_i, 2\chi.
\end{equation}
To fulfill this condition, one should assure the system remains in the eigenstate of the Hamiltonian $\tilde{H}$ all the time.

In summary, we have proposed an experimental scheme to realize a two-level PT-symmetric system with parameter being controllable through an embedding four-level cold atomic system. We further clarify the different properties between encircling and passing through EPs. And then we demonstrate that the change of the eigenstates and the relevant topological phase in the case of passing through EPs can be confirmed by measuring the phases of atomic levels and this novel phenomenon can be probed through the standard phase measurement. Our work proposes a method to realize PT-Symmetric Hamiltonian in clod atomic systems and therefore provides a powerful tool to explore the properties of PT-symmetry and PEs further.

\section*{Acknowledgements}

We thank Profs. Hui Yan and Shi-Liang Zhu and Dr. Feng Mei for helpful
discussions. This work was supported by the NKRDP of China (Grant
No. 2016YFA0301803), the NSFC (Grants No. 11604103), the NSF of
Guangdong province (Grants No. 2015A030310277, No. 2016A030310462, and No. 2016A030313436), and the SRFYTSCNU (Grants
No. 15KJ15 and No. 15KJ16).

\end{document}